\documentclass[12pt]{iopart}
\usepackage{iopams}
\usepackage{graphicx}
\newcommand{\ket}[1]{{\left| #1 \right>}}
\newcommand{\bra}[1]{{\left< #1 \right|}}

\begin{document}

\title{Group-theoretical approach to study atomic motion in a laser field}

\author{S V Prants}
\address{Laboratory of Nonlinear Dynamical Systems,\\ 
Pacific Oceanological Institute of the
Russian Academy of Sciences,\\
43 Baltiiskaya St., 690041 Vladivostok, Russia,\\
www.dynalab.poi.dvo.ru}
\ead{prants@poi.dvo.ru}

\begin{abstract}
Group-theoretical approach is applied to study behavior of lossless two-level atoms  
in a standing-wave laser field. Due to the recoil effect, the internal and external 
atomic degrees of freedom become coupled. The internal dynamics is described 
quantum mechanically in terms of the $SU(2)$ group parameters. The evolution 
operator is found in an explicit way after solving a single ODE for one of the 
group parameters. The translational motion in a standing wave is governed by 
the classical Hamilton equations which are coupled to the $SU(2)$ group equations. 
It is shown that the full set of equations may be chaotic in some ranges of 
the control parameters and initial conditions. It means physically that 
there are regimes of motion with chaotic center-of-mass motion and irregular 
internal dynamics. It is established that the chaotic regime is specified 
by the character of oscillations of the group parameter characterizing the mean 
interaction energy between the atom and the laser field. It is shown that the 
effect of chaotic walking can be observed in a real experiment with cold atoms 
crossing a standing-wave laser field. 
\end{abstract}
\pacs{2.20.Sv, 03.65.Fd, 05.45.Mt, 37.10.Jk}

\maketitle

\section{Introduction}

In quantum physics the unitary time evolution of a driven quantum system 
is described by the evolution operator equation
\begin{equation}
\label{1.2}
i\hbar \frac d{dt}\hat U(t,t_0)=\hat H\left[ {\bf h}(t)\right]
\hat U(t,t_0),\quad \hat U(t_0,t_0)=\hat I, 
\end{equation}
where $\hat U_{{\bf h}}(t,t_0)$ is a time evolution operator, $\hat H$ is a Hamiltonian 
and ${\bf h}(t)$ is a vector-function of the system's control 
parameters. From the abstract point of view, the evolution equation 
(\ref{1.2}) can be regarded as a differential equation on the group of 
dynamical symmetry. By dynamical symmetry we mean simply 
that the Hamiltonian can be expressed as a linear combination of operators
belonging to a finite-dimensional Lie algebra with $n$ basic elements. 
The parameters,
$g_k\,, k =1,2,\ldots,n$, of the respective Lie group satisfy to a set of $n$
first-order ordinary differential nonlinear equations which depend only on the
structure of the algebra and on $c$-number coefficients of the system's
Hamiltonian or the other governing operator \cite{Wei,Steinberg,JPA,JMP}. Thus, the dynamical group itself may 
be considered as a dynamical system.                                              

The dynamical-symmetry and Lie-algebraic approach has been successfully 
applied to describe the
time evolution of numerous physical systems in different disciplines
extending from classical mechanics \cite{Mukunda}, 
classical optics \cite{Stoler,Draght,Manko,Lie}
and quantum mechanics \cite{Manko71,Dodonov,Dodonov86,Dattoli,Hioe} to physics of neutrino oscillations 
\cite{JETP93}. As to study of dynamics of laser driven atoms, this  
approach has been applied to get Lie algebraic solution of the Bloch equations 
\cite{PLA90,JETP90}. 

The evolution of an isolated quantum system is regular, and the overlap of any two different 
quantum state vectors is a constant in course of time. All the 
expectation values of the quantum variables evolve in a quasiperiodic way at most. It does 
no matter how complicated is a dynamical symmetry of the quantum system under 
consideration and the corresponding Lie algebra. On the other hand, it is 
well known that even simple classical systems may be unstable and demonstrate 
chaotic behavior \cite{Zas,C91}. Classical instability is usually defined as an exponential 
separation of two initially close trajectories in time with an asymptotic rate given by the maximal Lyapunov
exponent $\lambda$. Such a behavior is possible because of the continuity of the 
classical phase space where the system's states can be arbitrary close to 
each other. The trajectory concept is absent in quantum mechanics, and the 
quantum phase space is not continuous due to the Heisenberg uncertainty 
principle. Perfectly isolated quantum systems are unitary, and there can be 
no chaos in the sense of exponential instability even if their 
classical limits are chaotic. What is usually understood under 
quantum chaos is, in essence, the special features of the quantum unitary 
evolution of the system under consideration (no matter how complicated the 
evolution  is) in the region of its control parameter values
and/or initial conditions at which its classical analogue is chaotic 
\cite{Z81,Haake,Reich,Stockmann}. 
In fact, it is not a special quantum problem. Any type of propagating waves 
(electromagnetic, sound or others), satisfying to a linear wave equation 
(which is an analogue of the Scrodinger equation), has the same property. 
Wave chaos is the special features of the wave field in the region 
of control parameters and/or initial conditions at which its ray  
analogue is chaotic \cite{Chaos04,MPVZ}. Thus, the quantum (wave) chaos 
problem is partly the problem of quantum (wave)-classical (ray) correspondence.  

Let us describe briefly the interconnection between the dynamical symmetries 
and dynamical chaos in physics of the atom-field interaction.

\begin{enumerate}
\item
The simplest problem is dynamics of a two-level atom at rest in an external 
laser field. From the dynamical symmetry  
point of view, the $SU(2)$ group, generated by the corresponding Hamiltonian, is driven 
by an external force that is not considered to be a dynamical
system.  It is the case of an external driving. The problem has been 
studied in Ref.~\cite{JMP}
where it has been shown that the evolution of atomic internal variables in a 
linearly polarized bichromatic laser field with incommensurate frequencies 
may be very complicated on the Bloch sphere albeit regular. It is simply because 
the dynamics takes place on the two-dimensional surface of the Bloch sphere.

\item
If we deal with a two-level atom at rest in an ideal cavity and take into 
account the response of the 
atom to the cavity radiation field, the semiclassical evolution of the 
coupled atom-field system may be chaotic in the sense of exponential 
sensitivity to small variations in initial conditions and/or parameters 
\cite{BZT,Milonni,Feinberg,Fox,Alekseev,JETPL97,PRE99}. 
This is 
the case of so-called dynamical driving \cite{JMP} when the $SU(2)$ 
group, generated by the atomic Hamiltonian, is driven by another dynamical system, 
the field one. We  deal now with 
a quantum system, the atom, which is coupled with a classical system, the 
radiation field governed by the Maxwell equations. The resulting 
Maxwell-Scrodinger (Bloch) equations constitute the five-dimensional set 
of nonlinear ordinary differential equations with two integrals of motion, 
the total atom-field energy (which is a constant in the absence of any losses) 
and the length of the Bloch 
vector. The motion in the phase space takes place on a three-dimensional 
manifold and may be chaotic due to transverse intersection of stable and unstable 
manifolds of hyperbolic points in some ranges of the control parameters, 
the values of the maximal Rabi frequency and the coupling strength 
\cite{PRE99}.  

\item
If a two-level atom moves within a standing-wave laser field in an open space,  
not in a cavity, the field may be considered as an external driving but 
one needs to take into account the atomic recoil effect, i.e. changes
of the atomic momentum after absorption or emission photons. If the atom is 
not especially cold, we may treat its translation degree of freedom classically. 
It is again the case of the dynamical driving with the $SU(2)$ 
group driven by the dynamical system which is now the classical atomic 
degree of freedom. The governing 
Hamilton-Scrodinger equations constitute the five-dimensional set 
of nonlinear ordinary differential equations with two integrals of motion, 
the atomic total energy, including the kinetic one, and the length of the Bloch 
vector. The motion in the phase space takes place on a three-dimensional 
manifold and may be chaotic in some ranges of the control parameters, 
the values of the maximal Rabi and atomic recoil frequencies.
A number of nonlinear dynamical
effects have been analytically and numerically demonstrated
with this system: chaotic Rabi oscillations 
 \cite{JETPL01,JETPL02}, Hamiltonian chaotic atomic
transport and dynamical fractals \cite{JETP03,PLA03,PRA07}, 
L\'evy flights and anomalous diffusion
 \cite{PRA02,JETPL02}. These effects are caused by
local instability of the center-of-mass motion in a laser
field. A set of atomic trajectories under certain conditions becomes 
exponentially sensitive to small
variations in initial quantum internal and classical external states
or/and in the control parameters, mainly, the atom-laser detuning. 
\end{enumerate}

In this paper we consider the physical situation mentioned in the third part 
of our nomenclature to focus at the ultimate reasons of chaotic atomic external 
and internal motion and its connection with the $SU(2)$ dynamical symmetry.

\section{Lie algebraic solution for the evolution operator with the $SU(2)$ 
dynamical symmetry}

In a variety of physical problems $SU(2)$ appears to be a group of dynamical
symmetry. It is known \cite{Wei,JMP} that the set of three equations for the 
$SU(2)$ group parameters can be reduced to a single second-order
differential equation. The form of this governing equation depends on the
choice of the basis and its exponential ordering. The appropriate choice of
parameterization of the dynamical group is especially important if we need 
to solve explicitly the governing equation for a given physical Hamiltonian.

The Hermitian Hamiltonian of a quantum system with the $SU(2)$ dynamical
symmetry can be cast in the general form 
\begin{equation}
\label{3.1}
\hat H(t)=h_0(t)\hat R_0+h^{*}(t)\hat R_{-}+h(t)\hat R_{+}, 
\end{equation}
where $\hat R_0$ and $\hat R_{\pm }$ are the generators that satisfy the commutation
relations 
\begin{equation}
\label{2.9}
\left[\hat R_{-},\hat R_{+}\right] =-2\hat R_0,\quad \left[\hat R_0,\hat R_{\pm
}\right] =\pm \hat R_{\pm }.
\end{equation}
It is convenient to choose the following noncanonical 
parameterization of the $SU(2)$ group   
\begin{equation}
\label{4}
\hat U=\exp\Bigl[\Bigl(g_0-i\int\limits_0^th_0\,d\tau\Bigr)\hat R_0\Bigr]\;
\exp g_{-}\hat R_{-}\; \exp g_{+}\hat R_{+}\;.
\end{equation}
Substituting Eq.(\ref{4}) into Eq.(\ref{1.2}), one finds the set
of differential equations for the group parameters that can be reduced
to the single equation for the group parameter $g \equiv \exp(g_0/2)$
\begin{equation}
\label{5}\frac{d^2g}{dt^2}-\Bigl(\frac{{dh}/{dt}}{h}+ih_0\Bigr)\frac{dg}{dt}
+\mid h\mid ^2g=0\;,\; g(0)=1\;,\;\frac{dg}{dt}(0)=0\;.
\end{equation}
Once Eq.(\ref{5}) is solved analytically, all the other parameters in
the product (\ref{4}) may be expressed in terms of the parameter $g$ as
follows:
\begin{equation}
\label{6}g_{-}=\frac{ig({dg}/{dt})}{h}\exp\Bigl(-i\int\limits_0^th_0\,d\tau\Bigr)\;,\quad
\frac{{dg}_{+}}{dt}=-\frac {ih}{g^2}\exp\Bigl(i\int\limits_0^th_0\,d\tau\Bigr)\;.
\end{equation}
It is convenient to introduce the new variable 
\begin{equation}
\label{7}\tilde g\equiv g_{-}/g.
\end{equation}
Then any group element in the parameterization (\ref{4}) can be described
by a pair of complex numbers $g$ and $\tilde g$ obeying the condition
\begin{equation}
\label{8}\mid{g\mid}^2+\mid{\tilde g\mid}^2=1\;.
\end{equation}

It should be noted that all these formulas are valid within any representation
and within any realization of the $SU(2)$ group. It is well known that the unitary
irreducible representations of $SU(2)$ are characterized by half-integer and
integer numbers $j$. The dimensionality of the $j$th representation is equal to
$2j+1$. In the $(2j+1)$-dimensional space of representation there is
a canonical basis 
\begin{equation}
\label{9}\left| j,m\right\rangle ,\quad m=-j,-j+1,...,j\;. 
\end{equation}
The representation matrix elements in the noncanonical parameterization 
(\ref{4}) are given by \cite{JMP} 
\begin{equation}
\label{10}
\begin{array}{c}
U_{m^{\prime }m}^{(j)}=\exp \left[ -im^{\prime }\int\limits_0^th_0(\tau
)d\tau \right] \sum\limits_{l=-j}^j\left[
\frac{(j-m^{\prime })!(j-m)!}{(j+m^{\prime })!(j+m)!}\right] ^{1/2}\times
\\ \frac{(j+l)!}{(j-l)!(l-m)!(l-m^{\prime })!}g^{m+m^{\prime }}({\tilde g})^
{l-m^{\prime }}(-{\tilde g^{*}})^{l-m}\;. 
\end{array}
\end{equation}

To analyze the dynamics of a two-level quantum system we need the two-dimensional 
representation of the $SU(2)$ group. In this case the generators 
$R$'s are connected with the familiar Pauli matrices  
\begin{equation}
\label{19}\hat R_0=\frac 12 \hat \sigma_z=\frac 12\left|
\begin{array}{cc}
1 & 0 \\
0 & -1
\end{array}
\right| ,\quad \hat R_{-}=\hat \sigma_{-}=\left|
\begin{array}{cc}
0 & 0 \\
1 & 0
\end{array}
\right| ,\quad \hat R_{+}=\hat \sigma_{+}=\left|
\begin{array}{cc}
0 & 1 \\
0 & 0
\end{array}
\right| \;, 
\end{equation}
where 
\begin{equation}
\label{20}[\hat \sigma_{+},\,\hat \sigma_{-}]=\hat \sigma_z\;,
\quad[\hat \sigma_z,\,\hat \sigma_{\pm }]=\pm 2\hat \sigma_{\pm }\;.
\end{equation}
In this representation the Hamiltonian of a driven two-level system 
has the form 
\begin{equation}
\label{21} 
\hat H(t)=\frac 12 \hbar \omega _a\hat \sigma_z+\hbar \Omega ^{*}(t)\hat \sigma_{-}
+\hbar \Omega (t) \hat \sigma_{+} \;,
\end{equation}
where $\Omega (t)$ is a time-dependent function which is, in general, 
a complex-valued one. The temporal 
evolution of the two-level system is now governed by the equation 
\begin{equation}
\label{22}
\frac{d^2g}{dt^2}-\Bigl(\frac{{d\Omega}/{dt}}{\Omega}+i\omega_a
\Bigr)\frac{dg}{dt}+\mid \Omega\mid ^2g=0\;,\; g(0)=1\;,\;\frac{dg}{dt}(0)=0\;.
\end{equation}
The evolution matrix in the basis 
\begin{equation} 
\label{23} 
\ket{1}=\ket{\frac 12\;,-\frac 12}, \ket{2}= \ket{\frac 12\;,\frac 12}  
\end{equation} 
is given by 
\begin{equation}
\label{Uatom}
\hat U^{(1/2)}=\left(
\begin{array}{cc}
e^{-i\omega _a t/2} & 0 \\
0 & e^{i\omega _a t/2}
\end{array}
\right) \left(
\begin{array}{cc}
g & -
{\tilde g}^{*} \\ {\tilde g} & g^{*}
\end{array}
\right)\;.
\end{equation}

\section{The $SU(2)$ group--Hamilton equations for a two-level atom 
moving in a standing-wave laser field}

We consider a two-level atom with mass $m_a$ and transition
frequency $\omega_a$, moving with the momentum $P$ along the axis $X$
in a one-dimensional classical laser standing wave with the frequency $\omega_f$
and the wave vector $k_f$. In the frame,
rotating with the frequency $\omega_f$, the model Hamiltonian is
the following:
\begin{equation}
\hat H=\frac{P^2}{2m_a}+\frac{1}{2}\hbar(\omega_a-\omega_f)\hat\sigma_z-
\hbar \Omega_0\left(\hat\sigma_-+\hat\sigma_+\right)\cos{k_f X},
\label{Jaynes-Cum}
\end{equation}
where $\Omega_0$ is the maximal Rabi frequency which is proportional to 
the square root of the number of photons in the wave.
The laser field is assumed to be strong enough,
so we can treat the field classically.

In the process of emitting and
absorbing photons, atoms not only change their internal electronic states
but their external translational states change as well due to the photon
recoil. If the atomic mean momentum is large as compared to the photon momentum
$\hbar k_f$, one can describe the translational degree
of freedom classically. The position and momentum of a point-like atom
satisfy classical Hamilton equations of motion 
which we represent in the normalized form 
\begin{equation}
\dot x=\omega_r p,\quad \dot p=- <\hat\sigma_-(t)+\hat\sigma_+(t)>\sin x, 
\label{Hameqs}                                       
\end{equation}
where $x\equiv k_f X$ and $p\equiv P/\hbar k_f$ are normalized classical
atomic center-of-mass position and momentum, respectively.  
The dot denotes differentiation with respect to the dimensionless 
time $\tau\equiv \Omega_0 t$ and $\omega_r\equiv\hbar k_f^2/m_a\Omega_0\ll 1$ is 
the normalized recoil frequency. To compute the 
quantum expectation value $<\hat\sigma_-(t)+\hat\sigma_+(t)>$ we need to use 
the solution for the evolution operator (\ref{Uatom}). Supposing that 
the atom is initially in the ground state $\ket{1}$, we get 
\begin{equation}
<\hat\sigma_-(t)+\hat\sigma_+(t)>= \bra{1} \hat U^{\dagger}(t)\hat U(t)\ket{1}
=-(gG^{*}+g^{*}G),                     
\label{expvalue}
\end{equation}
where we introduce for convenience the new complex-valued variable
\begin{equation}
G \equiv -\frac{i \dot g^{*}}{\cos x}.
\label{G}
\end{equation}

The internal atomic dynamics is governed by Eq.~(\ref{22}) that can 
be rewritten in the form of two first-order equations for the 
complex-valued group parameters $g$ and $G$. The self-consistent set 
of equations for the coupled external and internal atomic degrees of 
freedom now reads as 
\begin{equation}
\dot x=\omega_r p,\quad \dot p=(gG^{*}+g^{*}G)\sin x, 
\\
\dot g=iG \cos x, \quad \dot G=-i\Delta G + ig \cos x,
\label{mainsys}
\end{equation}
where the normalized recoil frequency $\omega_r$ and the atom-field 
detuning, $\Delta\equiv(\omega_f-\omega_a)/
\Omega_0$, are the control parameters. 
The six-dimensional dynamical system (\ref{mainsys}) has two independent integrals of motion, 
the total energy,
\begin{equation}
H\equiv\frac{\omega_r}{2}p^2+(gG^{*}+g^{*}G)\cos x-
\frac{\Delta}{2}(GG^{*}-gg^{*}),
\label{H}
\end{equation}
and the integral, 
\begin{equation}
\label{gG}
\mid{g\mid}^2+\mid{G\mid}^2=1
\end{equation}
reflecting conservation of the norm of the atomic wave function. 
It is evident from the second integral (\ref{gG}) 
that the squared absolute values of the $SU(2)$ group parameters,  
$\mid{g\mid}^2$ and $\mid{G\mid}^2$, have the sense of the probability amplitudes 
to find the atom in the ground and excited states, respectively.  

The equations of motion (\ref{mainsys}) describe the mixed quantum-classical 
dynamics of a two-level atom in a one-dimensional standing-wave laser field. 
The dynamical $SU(2)$  group is responsible for internal atomic dynamics 
caused by the interaction of the atomic electric dipole moment with the strength 
of the electric component of the field. The quantum expectation value of the 
corresponding interaction energy is given by the combination of the 
$SU(2)$  group parameters (\ref{expvalue}). The classical translational degree 
of freedom is described by the Hamilton equations (see the first two equations 
in the set (\ref{mainsys})) governed by the interaction energy. In Introduction 
we called such a situation as a dynamical driving when the $SU(2)$ 
group, generated by the atomic quantum Hamiltonian, is driven by another 
dynamical system, the classical atomic degree of freedom. In fact, we deal  
not with a fully quantum system but with a quantum-classical hybrid which 
is described by the c-number nonlinear dynamical system (\ref{mainsys}) 
that may be chaotic in the strict sense of this term in some ranges of 
the control parameters and/or initial conditions.

\section{Dynamical chaos in the group-theoretical picture}

Equations (\ref{mainsys}) constitute a nonlinear 
autonomous dynamical system with three degrees of freedom and, in general, with the two 
integrals of motion, (\ref{H}) and (\ref{gG}). Thus, the dynamical system (\ref{mainsys}) may be chaotic 
in the sense of exponential sensitivity to small variation in initial 
conditions and/or the control parameters, $\omega_r$ and $\Delta$. The common test 
to confirm that is to compute 
the maximum Lyapunov exponent characterizing the mean rate of 
exponential divergence of initially close trajectories 
which serves as a quantitative measure of dynamical chaos \cite{Oseledetz,Pesin}: 
\begin{equation}
\label{5.11}\lambda ({\bf Q}_0,\Delta {\bf q}_0)=
\lim_{t\to \infty, \Delta {\bf q}_0 \to 0} \frac 1t\ln 
\frac{\left\| \Delta {\bf q}({\bf Q}_0,t)\right\| }{%
\left\| \Delta {\bf q}_0\right\|},
\end{equation}
where $\Delta {\bf q}$ is the vector in the phase space with the
components $\left\{ \Delta q_j,j=1,...,N\right\} $ and the norm $\left\|
\Delta {\bf q}\right\| $. In Eq.(\ref{5.11}), $\Delta {\bf q}_0$ and $%
\Delta {\bf q}({\bf Q}_0,t)$ denote the separation between two initially
adjacent trajectories at the initial moment $t=0$ and at time $t$,
respectively, ${\bf Q}_0$ is the initial position. 
If, at least, one of the Lyapunov exponents of the dynamical system under 
question is positive, then 
trajectories, starting close together in the phase space, separate 
exponentially as
time grows. This very sensitive dependence on initial conditions is one of
the main indicator of dynamical chaos.

\begin{figure}[htb]
\begin{center}
\includegraphics[width=0.45\textwidth,clip]{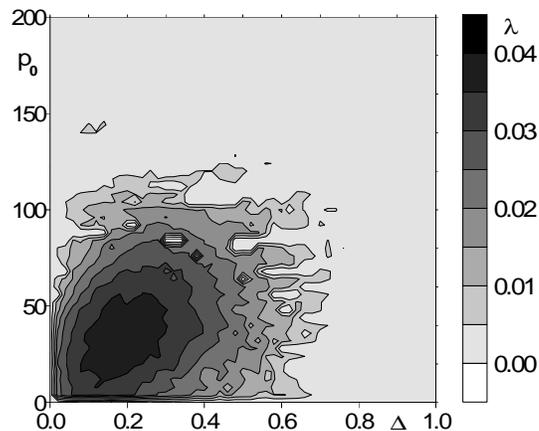}
\end{center}
\caption{Maximum Lyapunov exponent, $\lambda$, vs the atom-field detuning 
$\Delta$ (in units of the maximal Rabi frequency $\Omega$)
and the initial atomic momentum $p_0$  (in units of the 
photon momentum $\hbar k_f$) at $\omega_r=10^{-3}$.
Color codes the values of $\lambda$.}
\label{fig1}
\end{figure}

The result of computation
of the maximum Lyapunov exponent with the equations of motion 
(\ref{mainsys}) at $\omega_r=10^{-3}$ in dependence on
the detuning $\Delta$ and the initial atomic momentum  
$p_0$ is shown in Fig.~\ref{fig1}. 
Color in the plot codes the value of the maximum Lyapunov exponent $\lambda$.
In white regions in Fig.~\ref{fig1} the values of $\lambda$ are almost zero, 
and the atomic motion is regular in the corresponding ranges 
of $\Delta$ and $p_0$. In
shadowed regions positive values of $\lambda$ imply unstable motion.
The atoms with zero $\lambda$'s either oscillate in a regular way in a well 
of the optical potential or move ballistically over the hills of the potential 
with a  regular variation of their velocity.

\begin{figure}[htb]
\begin{center}
\includegraphics[width=0.45\textwidth,clip]{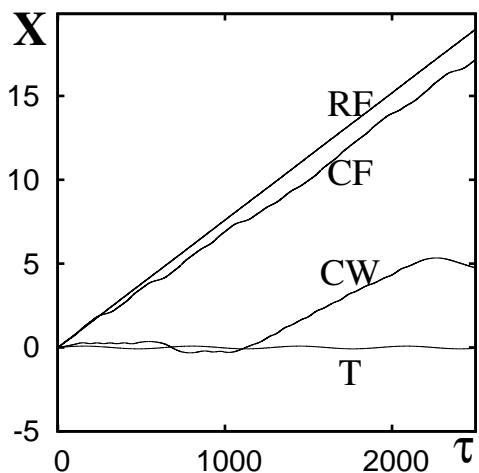}
\end{center}
\caption{Regimes of motion of two-level atoms in a 
one-dimensional deterministic standing-wave laser field. 
Trajectories in the real space at $\omega_r=10^{-3}$: 
regular flight (RF, $\Delta=0.8$, $p_0=45$), 
chaotic flight (CF, $\Delta=0.2$, 
$p_0=45$), chaotic walking (CW, $\Delta=0.2$, $p_0=10$) 
and trapping in a potential well (T, 
$\Delta=-0.2$, $p_0=5$). $x$ is in units of the wavelength $\lambda_f$.}
\label{fig2}
\end{figure}
\begin{figure}[htb]
\begin{center}
\includegraphics[width=0.4\textwidth,clip]{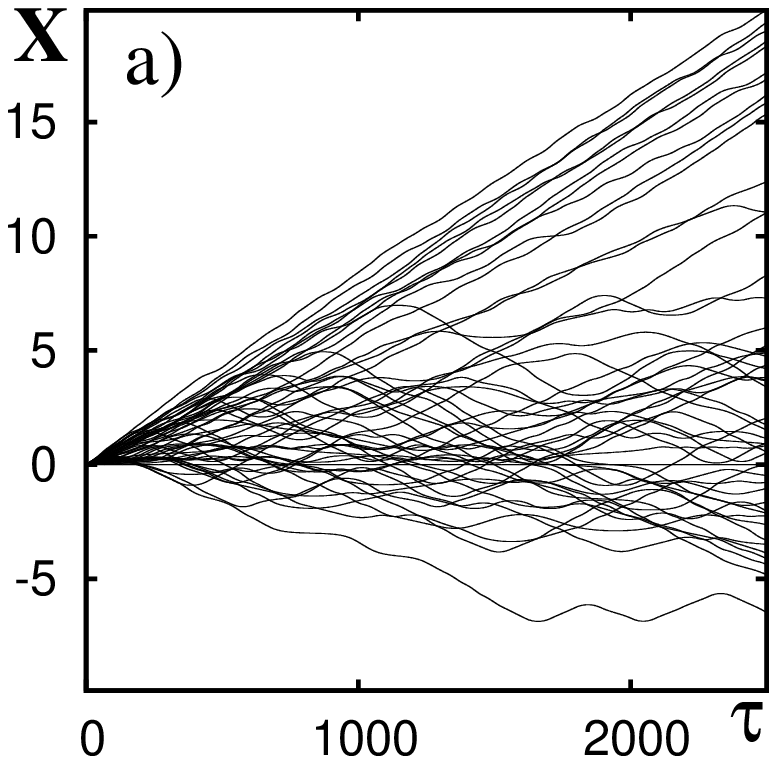}
\includegraphics[width=0.4\textwidth,clip]{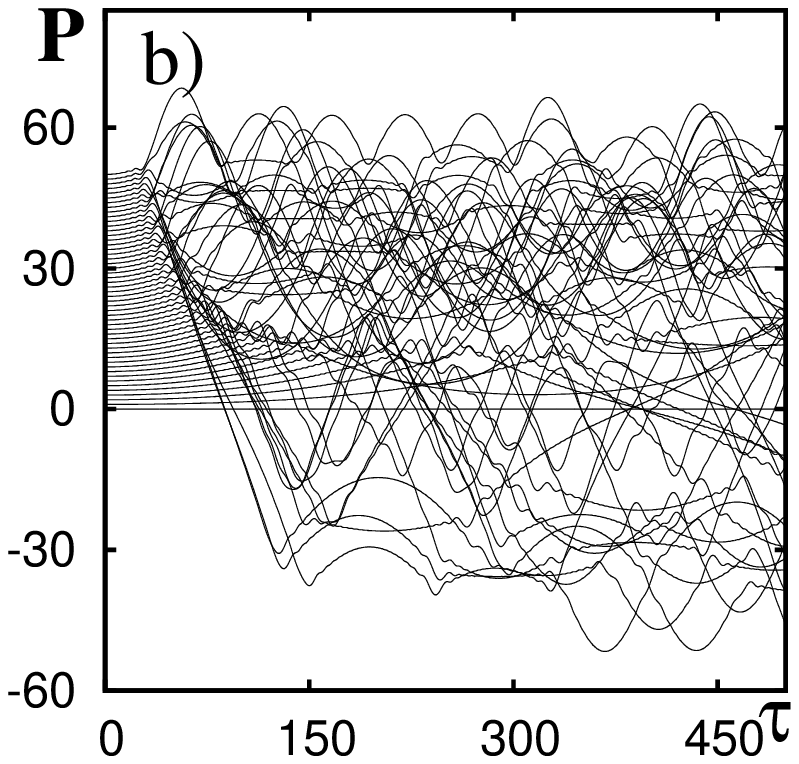}
\end{center}
\caption{Plots with 50 atomic trajectories 
with different values of the initial atomic momentum $0\le 
p_0 \le50$ but with the 
same initial position $x_0=0$ and the same other initial conditions. 
(a) Real space. (b) Momentum space.}
\label{fig3}
\end{figure}

At exact resonance, the equations of motion (\ref{mainsys}) 
become integrable due to an additional integral of motion, 
$gG^{*}+g^{*}G={\rm const}$, and we get $\lambda=0$. Thus at $\Delta=0$,
the center-of-mass motion and the motion in the space of the $SU(2)$ group 
parameters are regular. 

\begin{figure}[!htb]
\includegraphics[width=0.4\textwidth,clip]{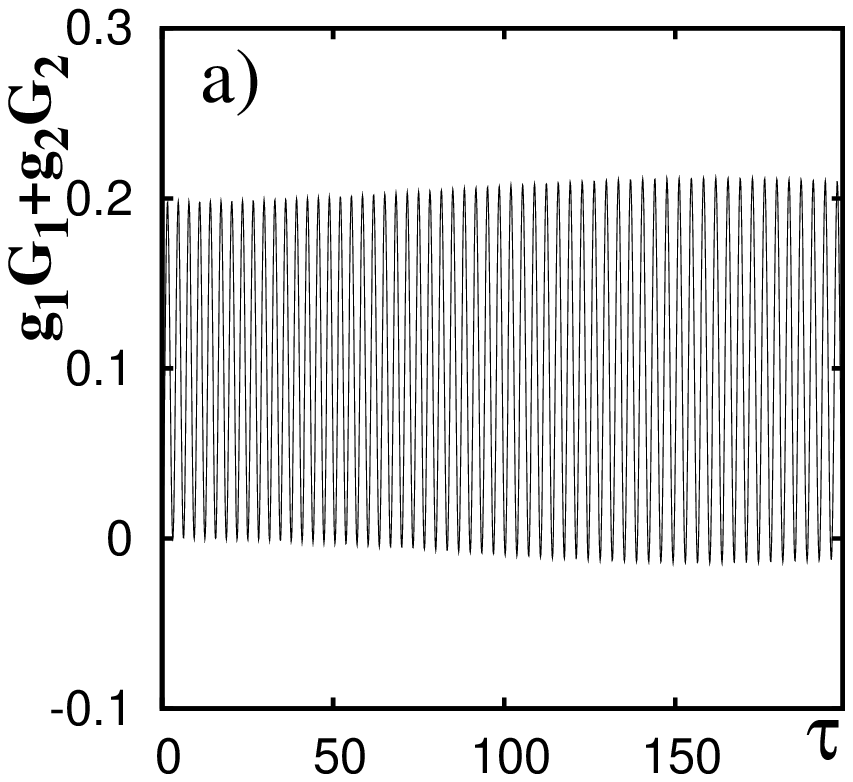}
\includegraphics[width=0.4\textwidth,clip]{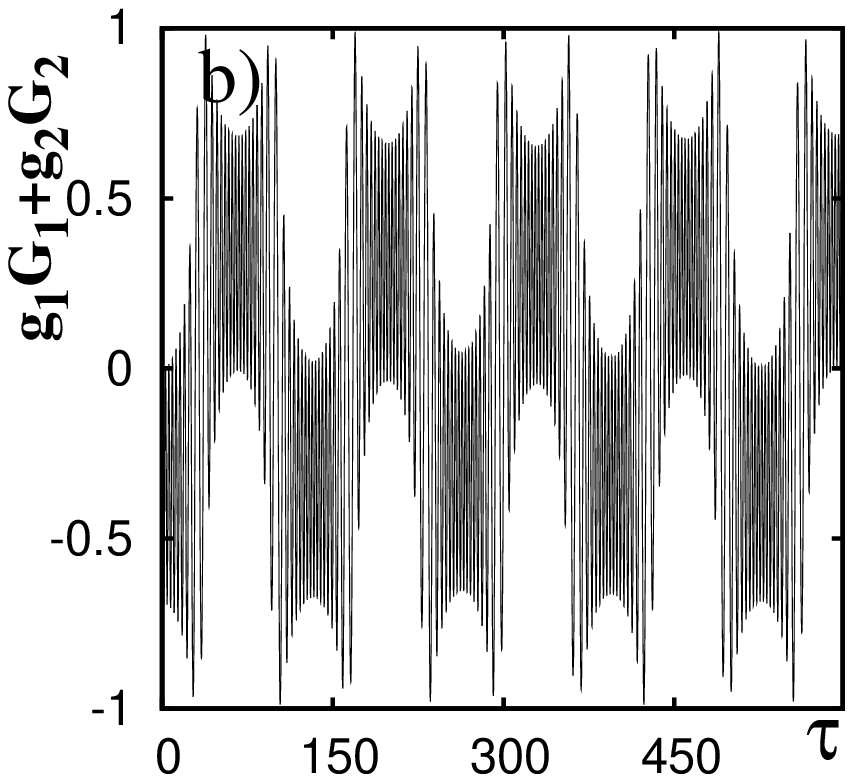}
\caption{Behavior of the mean atom-field interaction energy
$g_1G_1 + g_2G_2$  in the regimes of (a) regular oscillations 
in a well of the optical potential and (b) regular flight.}
\label{fig4}
\end{figure}
\begin{figure}[!htb]
\includegraphics[width=0.4\textwidth,clip]{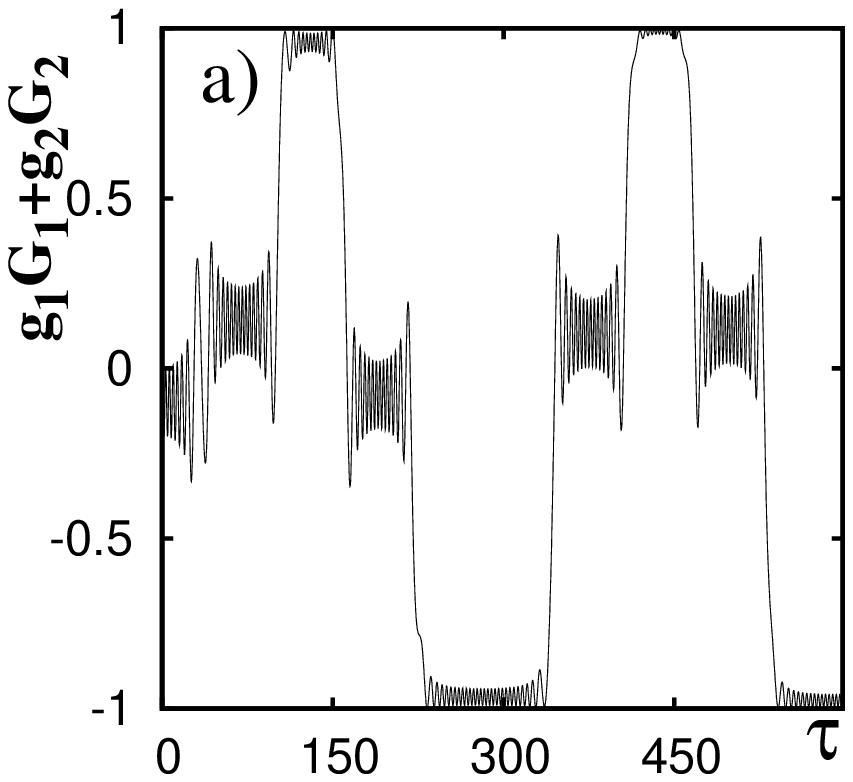}
\includegraphics[width=0.4\textwidth,clip]{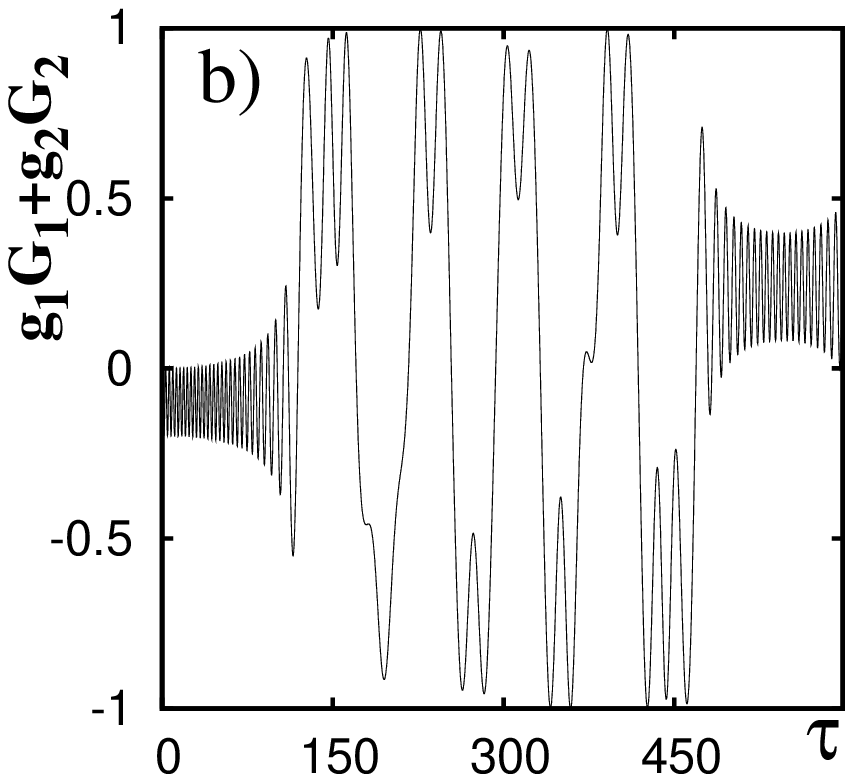}
\caption{The same as in Fig.~\ref{fig4} but in the regimes of 
(a) chaotic flight and (b) chaotic walking.}
\label{fig5}
\end{figure}

There are three possible chaotic types of motion of a two-level atom in a one-dimensional 
standing-wave laser field. In dependence on the initial conditions 
and the parameter values atoms may oscillate chaotically in a well of the 
optical potential, move ballistically over the hills of the potential 
with chaotic variations of their velocity or perform a chaotic walking.  
In the regime of the chaotic walking an atom {\it in a deterministic 
standing-wave field} alternates between flying through the standing-wave and 
being trapped in the wells of the optical potential. Moreover, it may change 
the direction of motion in a random-like way~\cite{PRA07}. 
We would like to stress that 
local instability produces chaotic center-of-mass 
motion in {\it a rigid} standing wave without any modulation of its 
parameters in difference from the situation with atoms in a periodically 
kicked optical lattice \cite{Raizen, Steck01, HH01}. 
To illustrate different types of motion we plot in Fig.~\ref{fig2} 
four trajectories of the atoms in the real space at $\omega_r=10^{-3}$ 
corresponding to a regular flight (RF), chaotic flight (CF), 
chaotic walking (CW) and trapping in a potential well (T).  
\begin{figure}[!htb]
\includegraphics[width=0.3\textwidth,clip]{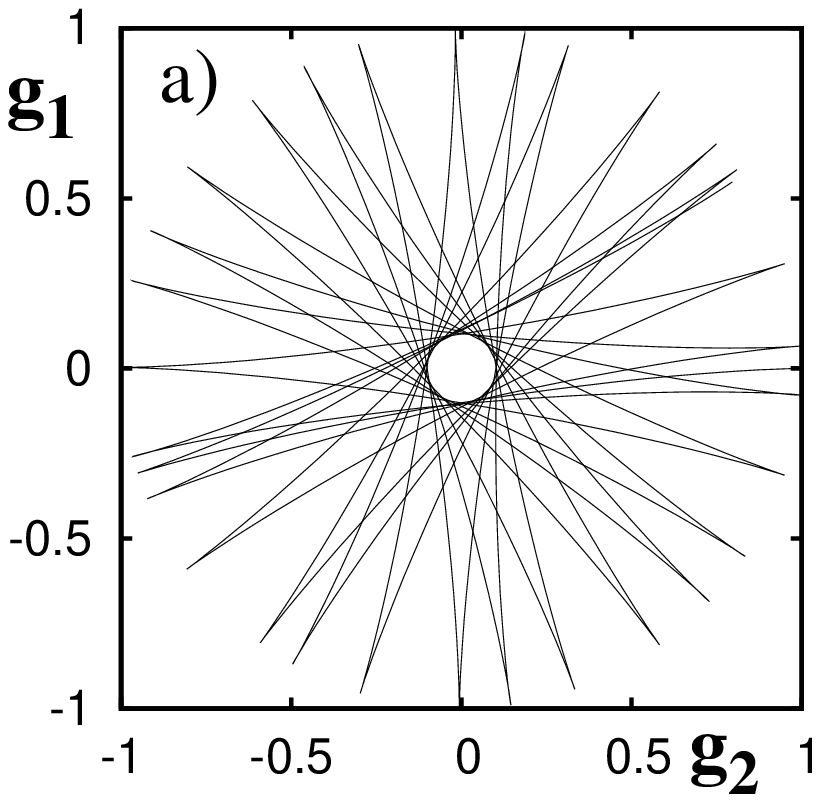}
\includegraphics[width=0.3\textwidth,clip]{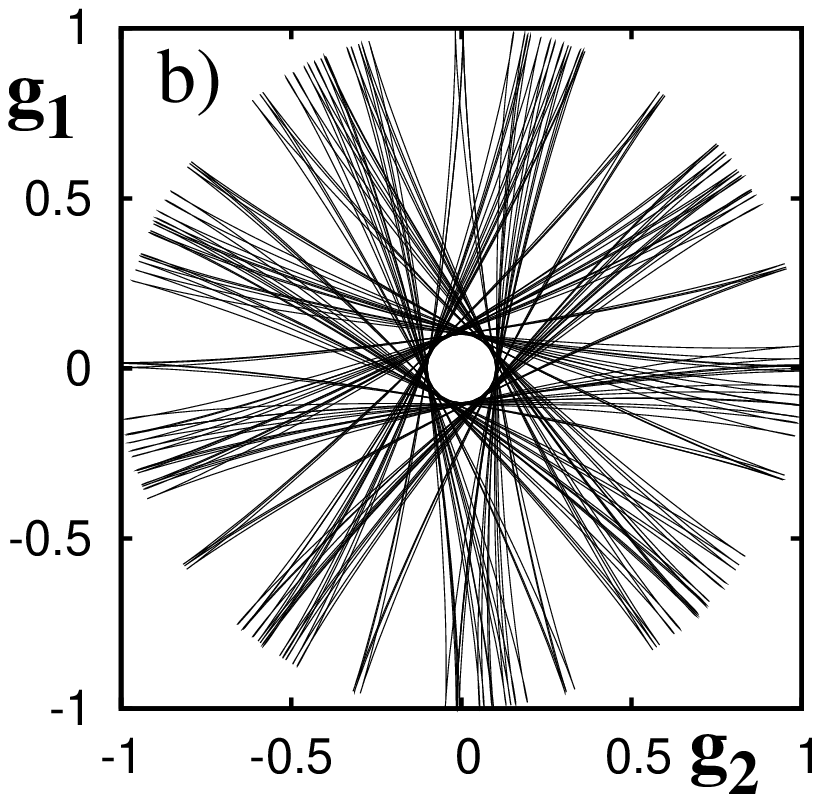}
\includegraphics[width=0.3\textwidth,clip]{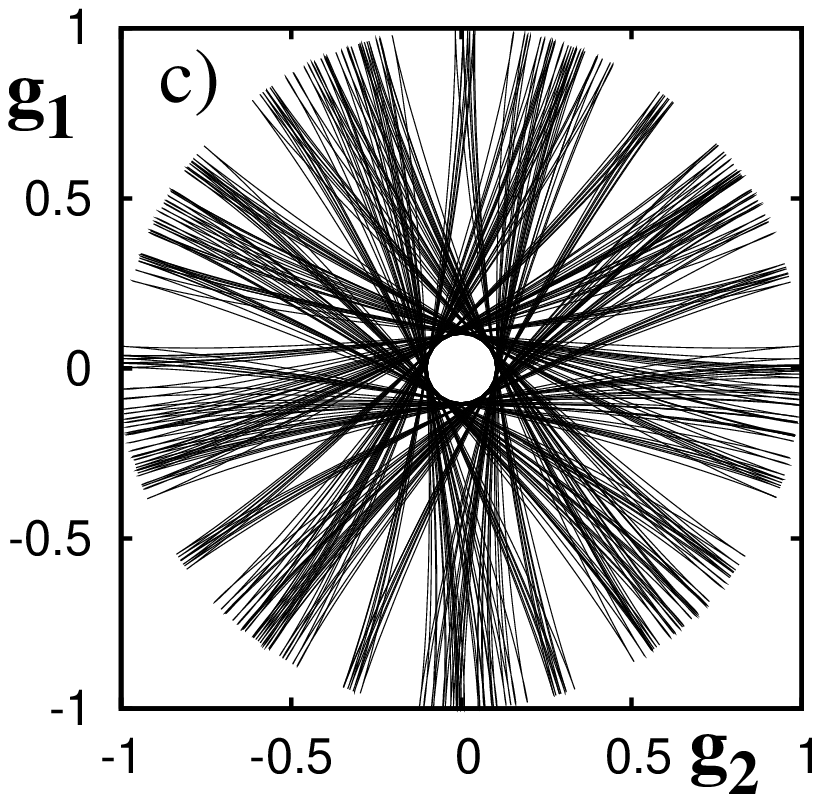}
\caption{Projections of the single trajectory of a trapped atom in the 
six-dimensional phase space on the plane of the complex-valued $SU(2)$ 
group parameter $g=g_1 +ig_2$ at 
(a) $\tau=100$, (b) $\tau=500$ and (c) $\tau=1000$.}
\label{fig6}
\end{figure}
\begin{figure}[!htb]
\includegraphics[width=0.3\textwidth,clip]{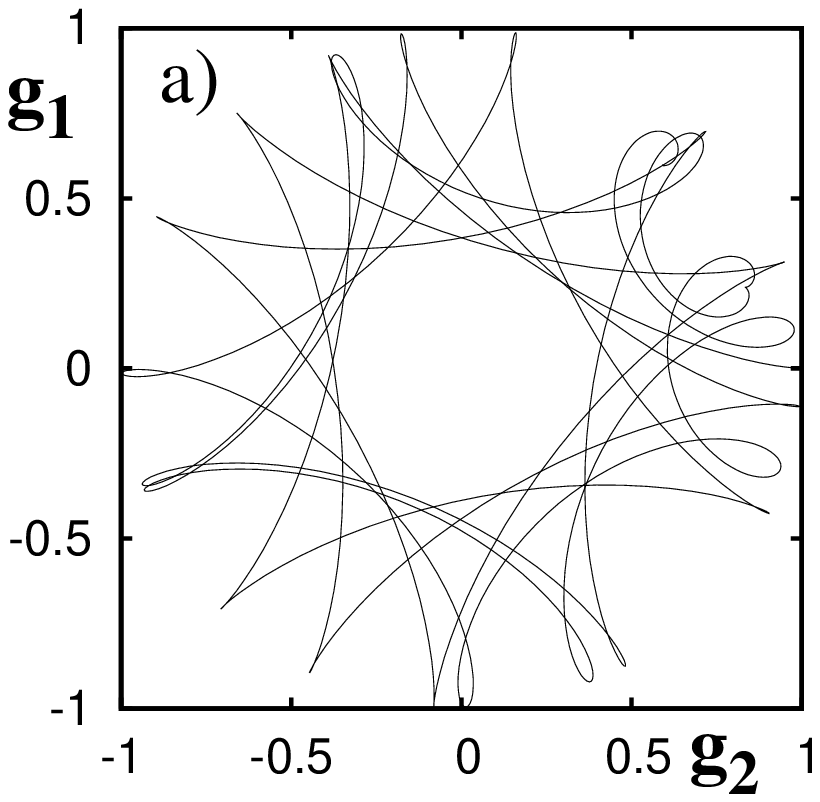}
\includegraphics[width=0.3\textwidth,clip]{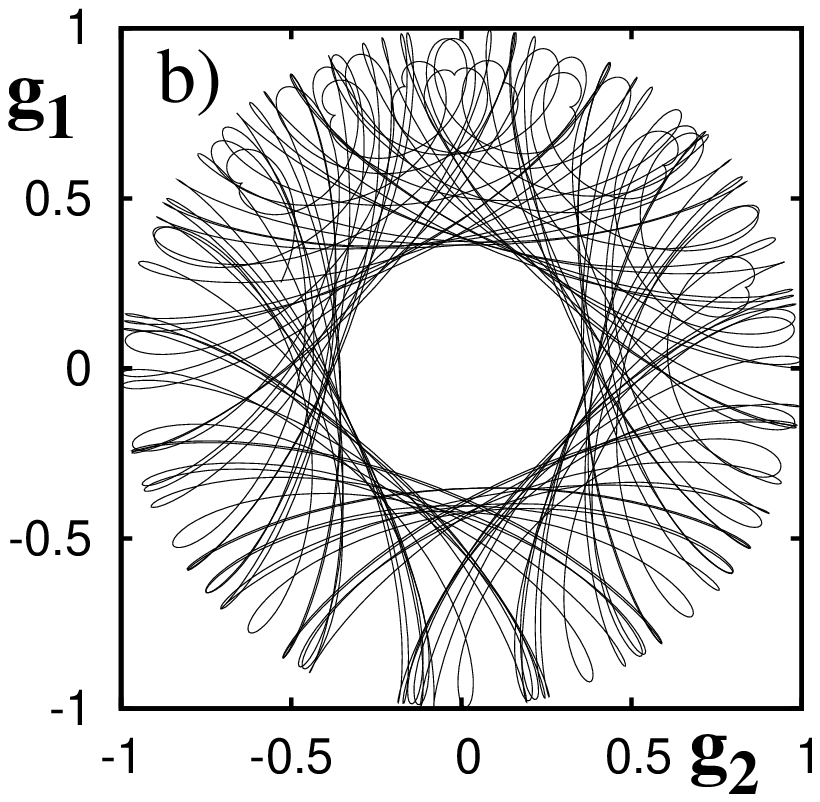}
\includegraphics[width=0.3\textwidth,clip]{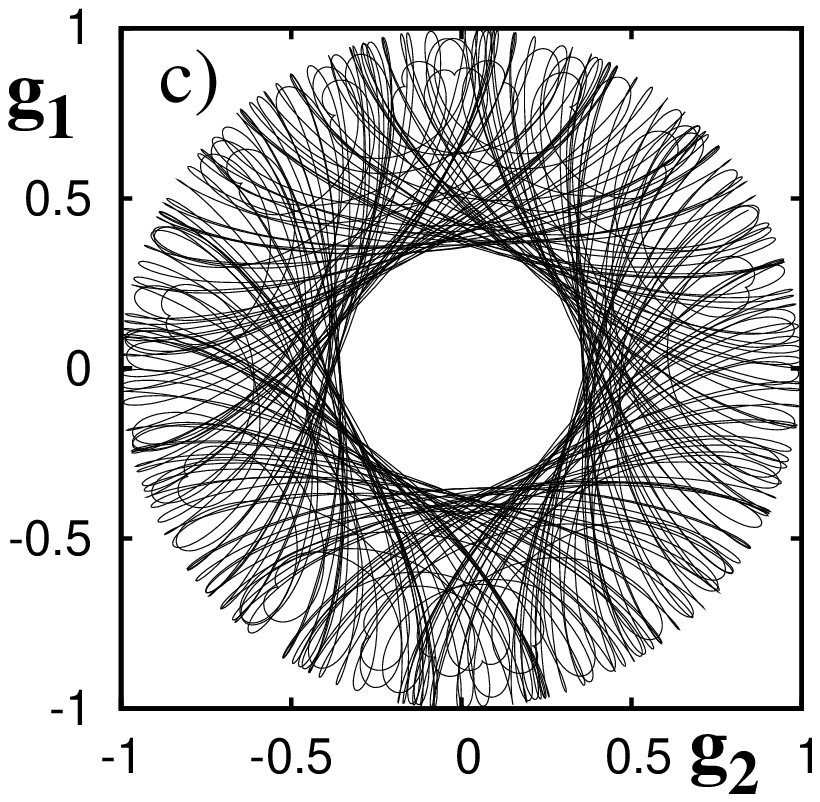}
\caption{The same as in Fig.~\ref{fig6} but for a regular flight.}
\label{fig7}
\end{figure}

Let us estimate the values of the control parameters and 
the initial conditions under which atoms oscillate in the first well of the optical 
potential, move ballistically or walk chaotically. At small detunings, 
$\Delta \ll 1$, 
the total energy (\ref{H}) consists of the kinetic one, $K=\omega_rp^2/2$, 
and the potential one,  $U=(gG^{*}+g^{*}G)\cos x=(g_1G_1 + g_2G_2)\cos x$, 
the sum of 
which is conserved approximately in course of time. The maximal absolute value of the 
optical potential energy is 1. 
Let the atom is prepared in the ground state $\ket{1}$, i.e., 
$g_1(\tau=0)=1$, $g_2(\tau=0)=G_1(\tau=0)=G_2(\tau=0)=0$ and $U_0=0$. 
If $K_0> |U_{\rm max}|=1$, then the atom will move 
ballistically. This occurs if the initial atomic momentum, $p_0$, 
satisfies to the condition $p_0> \sqrt{2/\omega_r}\simeq 44$ at 
$\omega_r=10^{-3}$. If the initial conditions are chosen to give 
$0\le H_0=K_0+U_0 \le 1$, then the atom performs a chaotic walking. This occurs 
at $0\le p_0\le 44$. The atom will be trapped 
in the first well of the optical potential if $H_0<0$. It is posiible with the 
initial conditions chosen only if $\Delta<0$.

To demonstrate strong dependence of the atomic motion on 
initial conditions we compute 50 trajectories 
with different values of the initial atomic momentum, $0 \le p_0 \le 50$, 
but with the 
same initial position, $x_0=0$, and the same other initial conditions. 
Figure~\ref{fig3} gives an impressive image of dynamical chaos with atoms 
in a laser field both in the real and momentum spaces.  
Most of the atoms in this bunch (with $0\le p_0 \le 44$)  walks chaotically, 
changing the direction of motion in course of time. 
Atomic trajectories with close initial conditions diverge 
in the real one-dimensional space in such a way that it is practically 
impossible to predict their final position after the predictability time
\begin{equation}
\tau_p \approx \frac{1}{\lambda} \ln \frac{\Delta x}{\Delta x(0)},
\label{predtime}
\end{equation}
where $\Delta x$ is the confidence interval and 
$\Delta x_(0)$ is the practically inevitable error in measuring the initial 
atomic position.

It follows from (\ref{mainsys}) that the translational motion 
is described by the equation for a nonlinear physical
pendulum with the frequency modulation
\begin{equation}
\ddot x - 2\omega_r  (g_1G_1 + g_2G_2)\sin x=0.
\label{xosc}
\end{equation}
It is clear that the regime of the center-of-mass  
motion is specified by the character of oscillations of the group parameter,  
$g_1G_1 + g_2G_2$, which has the sense of the mean interaction energy 
between the atom and the laser field (see the integral of motion (\ref{H})). 
If the atom is trapped in the first well of the optical potential, 
its center of mass oscillates between the first negative and positive nodes, 
$-\pi/2<x< \pi/2$. If, in addition, the control parameters are chosen in 
appropriate way, it will oscillate periodically. This case is shown in 
Fig.~\ref{fig4}~a ($\Delta=-0.2$, $p_0=5$) with periodic albeit modulated 
oscillations of the quantity $g_1G_1 + g_2G_2$. 
Figure~\ref{fig4}~b is plotted for another regime of the center-of-mass motion, 
a regular ballistic flight with $\Delta=0.8$ and $p_0=45$. The quantity 
$g_1G_1 + g_2G_2$ again oscillates periodically but with the modulation 
period that is equal to the flight time between two adjacent nodes of 
the laser standing wave, $T_f \simeq \pi/\omega_rp_0 \simeq 70$.

Behavior of the group parameter $g_1G_1 + g_2G_2$ is 
absolutely different in the chaotic regimes of motion, CF and CW.  
In the regime of chaotic ballistic flight  
(see Fig.~\ref{fig5}~a with $\Delta=0.2$ and $p_0=45$), shallow oscillations 
of that quantity are interrupted by jumps of different amplitudes that occur 
when the atom crosses each node of the standing wave. In the regime of chaotic 
center-of-mass walking (see Fig.~\ref{fig5}~b with $\Delta=0.2$ and $p_0=10$), 
oscillations of the quantity $g_1G_1 + g_2G_2$ look even more complicated.
We may conclude that namely the chaotic oscillations of the mean interaction energy 
between the atom and the laser field, $g_1G_1 + g_2G_2$, in some ranges of 
the control parameters, $\omega_r$ and $\Delta$, and initial atomic momentum 
$p_0$ are responsible for the chaotic center-of-mass motion. 
\begin{figure}[!htb]
\includegraphics[width=0.3\textwidth,clip]{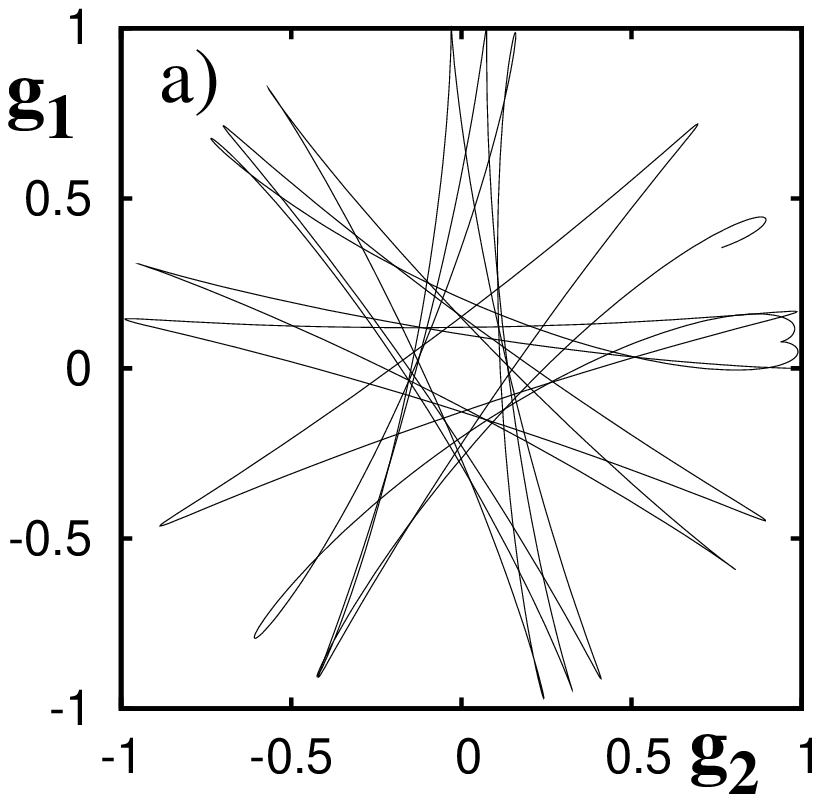}
\includegraphics[width=0.3\textwidth,clip]{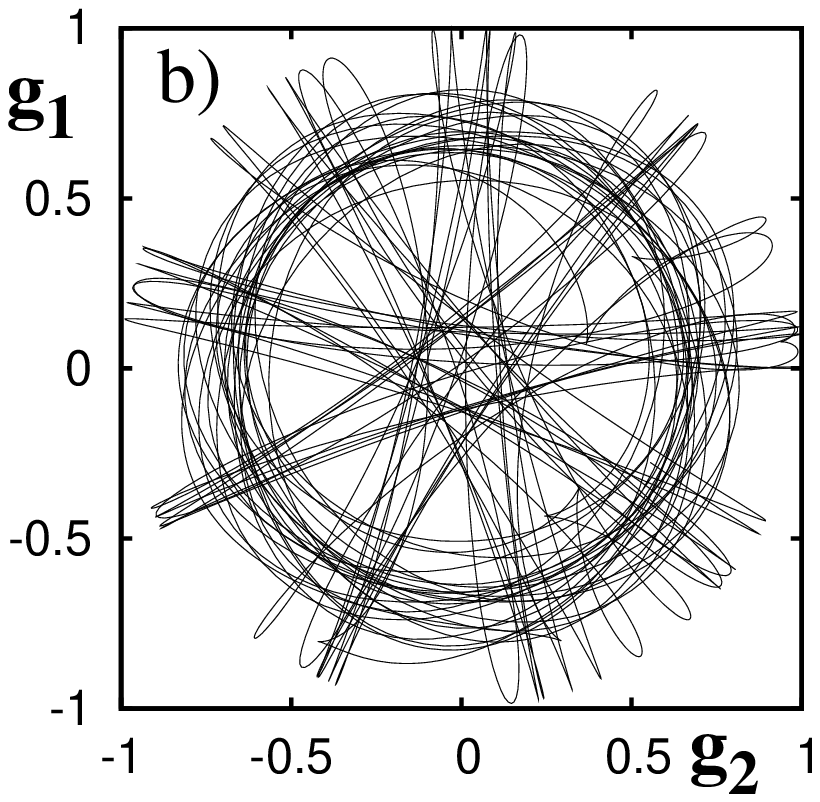}
\includegraphics[width=0.3\textwidth,clip]{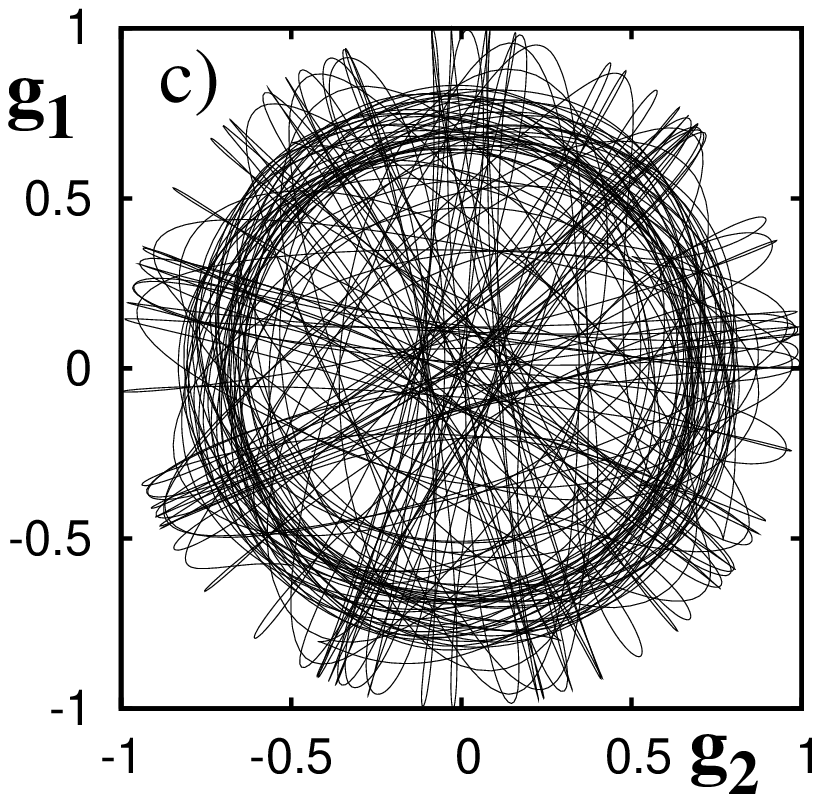}
\caption{The same as in Fig.~\ref{fig6} but for a chaotic flight.}
\label{fig8}
\end{figure}
\begin{figure}[!htb]
\includegraphics[width=0.3\textwidth,clip]{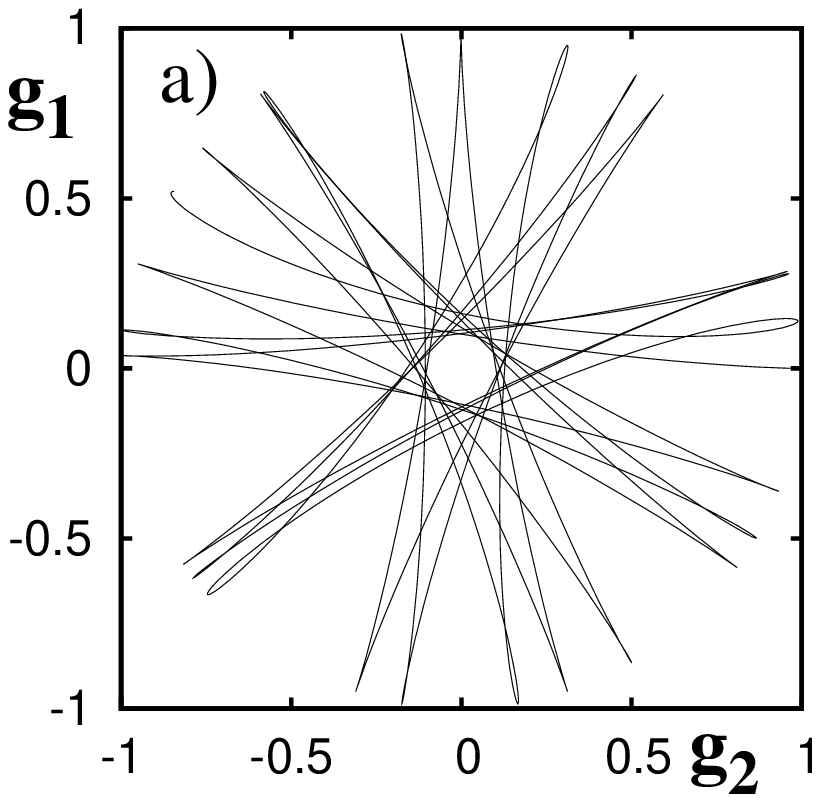}
\includegraphics[width=0.3\textwidth,clip]{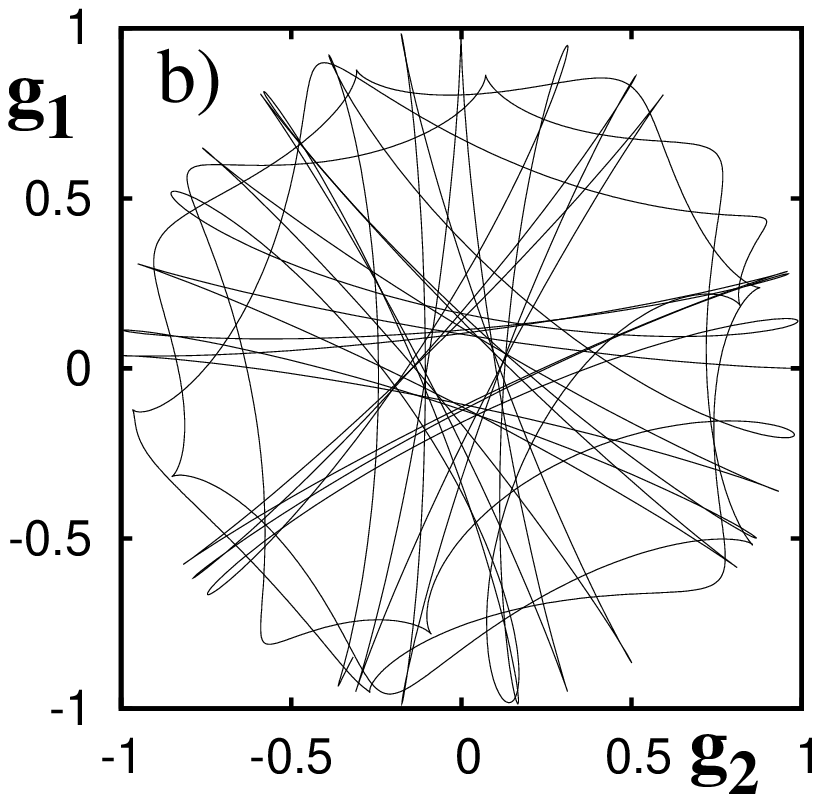}
\includegraphics[width=0.3\textwidth,clip]{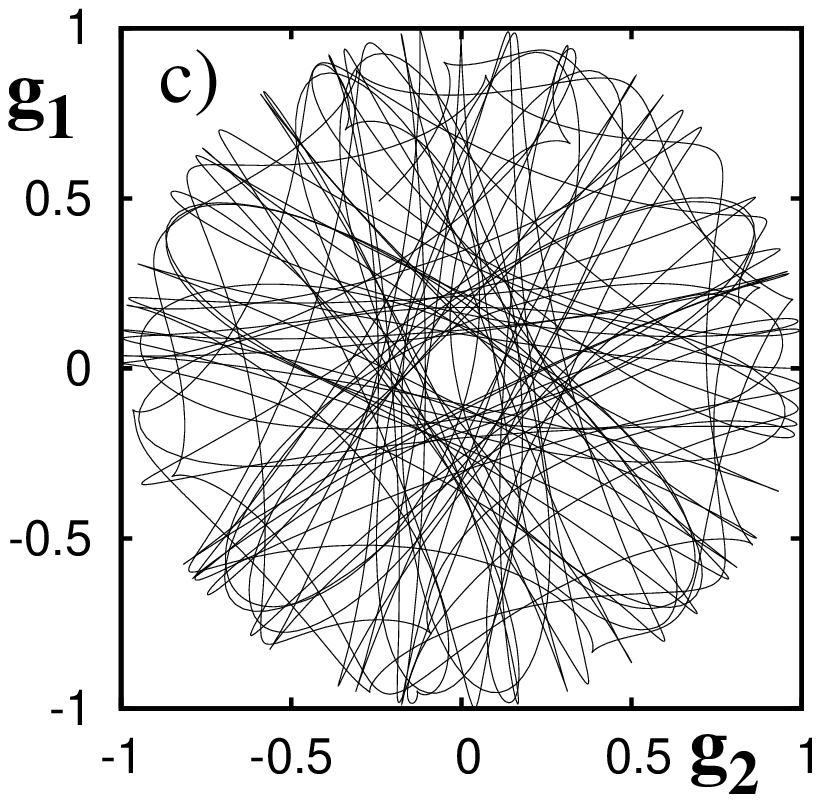}
\caption{The same as in Fig.~\ref{fig6} but for a chaotic walking.}
\label{fig9}
\end{figure}

The equations of motion (\ref{mainsys}) can be recast in the form of the 
two second-order differential equations, the classical one (\ref{xosc}), 
describing the center-of-mass motion, and the quantum one 
\begin{equation}
\ddot g + (i\Delta + \dot x \tan x)\dot g +g \cos^2x=0, 
\label{gosc}
\end{equation}
describing the internal atomic dynamic in terms of the complex-valued $SU(2)$ 
group parameter $g=g_1+ig_2$. In order to illustrate how different may be behavior 
of the quantum degree of freedom of the quantum-classical hybrid, 
we compute the evolution of the real and imaginary parts of $g$ in course of time. 
The results are shown in Figs.~\ref{fig6}-- \ref{fig9} with different regimes 
of motion. The plots give projections of the single atomic trajectory in the 
six-dimensional phase space on the plane of the complex-valued $SU(2)$ 
group parameter $g$ at the time moments $\tau=100$, $\tau=500$ and $\tau=1000$. 

The plots with a periodically oscillating atom in a trap (Fig.~\ref{fig6}) 
and with a regular flight (Fig.~\ref{fig7}) demonstrate the strictly periodic 
patterns on the $g_1$--$g_2$ plane with  forbidden regions in the center.  
Internal dynamics of the atoms in the chaotic center-of-mass regimes of motion, 
chaotic flight in Fig.~\ref{fig8} and chaotic walking in Fig.~\ref{fig9}, 
is much more complicated. In both the cases, the trajectories 
on the $g_1$--$g_2$ visit 
in course of time all the accessible part of the plane with $\mid g\mid <1$. 

\section{How to observe chaotic walking of atoms in a real 
experiment}

\begin{figure}[!htb]
\includegraphics[width=0.4\textwidth,clip]{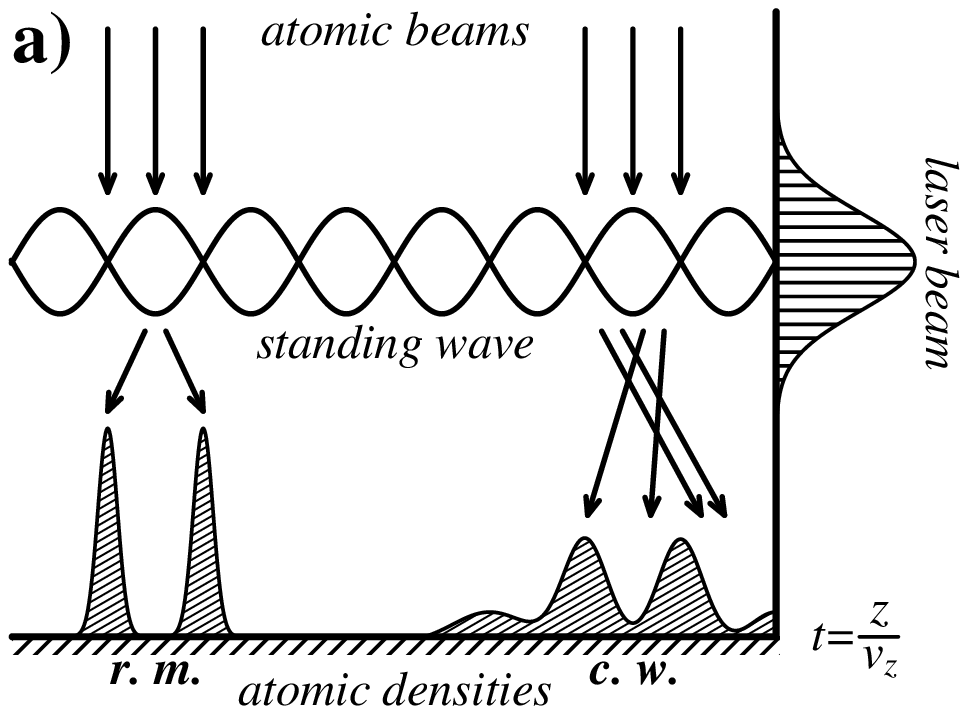}
\includegraphics[width=0.4\textwidth,clip]{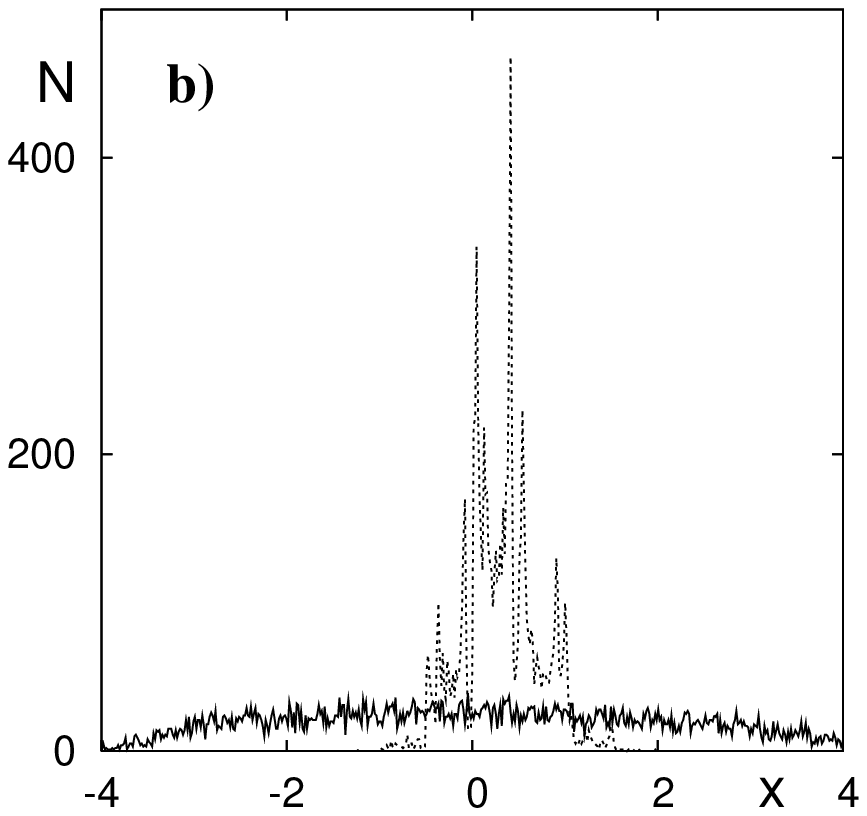}
\caption{(a) Scheme of the proposed experiment on 
observation of chaotic walking (cw) of atoms scattered 
at a Gaussian standing laser wave. 
(b) The distributions of $10^4$ lithium atoms at $\tau=1000$ 
($z=200$~microns) under the conditions of chaotic walking at $\Delta=0.2$ 
(bold curve) and regular motion (rm) at $\Delta=1$ (dashed curve).}
\label{fig10}
\end{figure}

In this section we propose the scheme of a real experiment 
to observe the effect of chaotic walking of atoms in a deterministic 
standing wave described in the previous section.
A beam of two-level atoms in the $z$ direction crosses  
a standing-wave laser field with optical axis in the $x$ direction 
(Fig.~\ref{fig10}a). One measures either the atomic density on 
a substrate as in the atom-lithography experiments \cite{Timp,Ober} 
or the spatial 
atomic distribution as in the atom optics experiments \cite{Raizen,Steck01,HH01}. 
In each type of the experiments the results are expected 
to be different in the regimes of chaotic atomic walking and regular motion.  
To switch between the regimes it is enough to vary the value of the detuning 
in the appropriate way.
The laser beam has the Gaussian profile
$\exp[-(z-z_0)^2/r^2]$ with $r$ being the $e^{-2}$ radius at the laser beam waist.
The longitudinal velocity of atoms, $v_z$,  
is much larger than their transversal velocity $v_x$ and is supposed to be 
constant. Therefore, the spatial laser profile may be replaced by the temporal one.

The Hamiltonian (\ref{Jaynes-Cum}) now takes the time-dependent form
\begin{equation}
\hat H=\frac{\hat P^2}{2m_a}+\frac{\hbar}{2}(\omega_a-\omega_f)\hat\sigma_z- \\
\hbar \Omega_0\exp[-(v_zt - \frac 32r)^2/r^2]
\left(\hat\sigma_-+\hat\sigma_+\right)\cos{k_f \hat X}
\label{Ham}
\end{equation}
with the same dynamical symmetry. Using the same normalization as before, 
we get the equations of motion 
\begin{eqnarray}
\ddot x - \omega_r \Omega(\tau) (gG^{*}+g^{*}G)\sin x=0\\
\ddot g + \left[i\Delta + \dot x \tan x - \frac{\dot \Omega(\tau)}
{\Omega(\tau)}\right]\dot g +g [\Omega(\tau)]^2\cos^2x=0 
\label{eqs}
\end{eqnarray}
with the time-dependent coefficient $\Omega(\tau)=\exp[-(\tau-\frac{3}{2} 
\sigma_{\tau})^2/\sigma^2_{\tau}]$, where $\sigma_{\tau}\equiv r\Omega_0/v_z$ 
is the normalized characteristic interaction time. 

To be concrete let us take lithium atoms with the relevant 
transition $2S_{1/2}-2P_{3/2}$, the corresponding wavelength
$\lambda_a = 670.7$~nm and the recoil frequency $\nu_{\rm rec}=63$~KHz. 
With the maximal Rabi frequency $\Omega_0/2\pi \simeq 126$~MHz 
and the radius of the laser beam $r = 0.05$~cm 
one gets $\omega_r = 10^{-3}$ and $\sigma_{\tau}=400$. 
To simulate a real experiment let us consider a beam of $10^4$ lithium atoms with 
the initial Gaussian position and momentum distributions   
(the rms $\sigma_x=\sigma_p=2$, the average values, $x_0=0$, 
and $p_0=10$) and compute their position distribution at a fixed moment 
of time. In Fig.~\ref{fig10}~b we compare the atomic 
position distributions at $\tau=1000$ ($z=200$~microns) for the chaotic 
walking at $\Delta =0.2$ (bold curve) and the regular motion at 
$\Delta =1$ (dashed curve). The difference is evident. 
In the chaotic regime atoms are distributed more or less homogeneously 
over a large 
distance of 8 wavelengths along the $x$-axis whereas in the regime 
of the regular motion they form a few peaks in a much more narrow interval. 
Thus, we predict that under the conditions of chaotic walking  
there should appear a less contrast and more broadened atomic relief as 
compared to the case of regular motion because a large number of atoms 
are expected to be deposited 
between the nodes as a result of chaotic walking along the standing-wave axis.

\section{Conclusion}

We have studied behavior of lossless two-level atoms  
in a one-dimensional standing-wave laser field in the group-theoretical picture. 
In this picture we have represented the internal quantum atomic dynamics 
in terms of the dynamical 
$SU(2)$ group parameters and the center-of-mass motion by the classical 
Hamilton equations. 
Thus, we have modeled the system by a quantum-classical hybrid with coupled 
quantum and 
classical degrees of freedom. We have derived the corresponding set of the 
$SU(2)$ group-Hamilton equations of motion with, in general, two integrals of motion. 
This set has been numerically shown to be chaotic in some ranges of the control parameters 
and initial conditions. We have found five different regimes of the 
center-of-mass motion including chaotic walking when  
an atom {\it in an absolutely deterministic 
standing-wave field} may change the direction of motion in a random-like way 
alternating between flying in the optical potential and 
being trapped in its wells. 
All the regimes have been illustrated by the trajectory plots in the real and 
momentum spaces. It has been established that the instability of motion and 
dynamical chaos are caused by the character of oscillations of the group 
parameter characterizing the mean interaction energy between the atom and the 
laser field. Projections of atomic trajectories in the 
six-dimensional phase space on the plane of the complex-valued $SU(2)$ 
group parameter $g$ have been shown to form regular and irregular patterns 
in the regimes of regular and chaotic center-of-mass motion, respectively. 

We proposed the scheme of an experiment on the scattering of atomic beams at a 
standing-wave laser field that could directly image chaotic walking of 
atoms along the optical axis. 
In a real experiment the final spatial distribution can be recorded via fluorescence or 
absorption imaging on a CCD, commonly used methods in atom optics experiments 
yielding information on the number of atoms and the cloud's spatial size. 
The other possibility is a nanofabrication 
where the atoms after the interaction with the standing wave are deposited on a silicon 
substrate in a high vacuum chamber. In this case the spatial distribution 
can be analyzed with an atomic force microscope. The modern tools of atom optics enable to create 
narrow initial atomic distributions in position and momentum, reduce coupling 
to the environment and technical noise, create one-dimensional optical 
potentials, and to measure spatial and momentum distributions 
with high sensitivity and accuracy \cite{Raizen,Steck01,HH01}.

The results obtained can be applied to other models of the atom-field 
interaction as well. In particular, relaxation processes in two-level atoms 
can be described 
within the framework of the $SO(3)$ dynamical-symmetry approach to solving 
the Bloch equations \cite{JETP90}. 
Moreover, one may consider by the method developed in this paper 
the dynamics not only of two-level atoms but of three-, four- and multilevel atoms 
excited by a few laser fields at different atomic transitions. 
If the corresponding model  
Hamiltonian has the $SU(2)$ dynamical symmetry, then one may use the 
solution obtained in Sec.~2 that is 
valid for any representation of the $SU(2)$ group. 

The model considered can be generalized to the case with two-level atoms  
inside a high-quality cavity with a quantized field. 
In the rotating wave approximation the state space of the 
corresponding Jaynes-Cummings model splits up into an infinite class 
of two-dimensional non-communicating
subspaces each of which being labeled by eigenvalues of the Casimir operator. 
The system evolves in such a way that transitions between the 
subspaces with different eigenvalues are forbidden. The 
solution for the time-evolution operator in each of these subspaces is
given by the matrix (\ref{23}) with the group parameter satisfying to 
the equation similar to (\ref{5}). The resulting equations of motion for 
the coupled atom-field system are 
expected to constitute an infinite-dimensional set of the type (\ref{mainsys}) 
with the group equation (\ref{23}) acting in each 
of the subspaces labeled by its own  eigenvalue. This set is expected to 
admit very different regimes of motion including chaotic ones.

\section*{Acknowledgments}
This work was supported  by the Russian Foundation for Basic Research
(project no. 09-02-00358), by the Integration grant from the Far-Eastern 
and Siberian branches of the Russian Academy of Sciences and by the Program
``Fundamental Problems of  Nonlinear Dynamics'' of the Russian
Academy of Sciences. 

\end{document}